\documentclass[acmsmall,screen]{acmart}

\acmBooktitle{Companion Proceedings of the 34th ACM Symposium on the Foundations of Software Engineering (FSE '25), June 23--27, 2026, Trondheim, Norway}

\AtBeginDocument{%
  \providecommand\BibTeX{{%
    \normalfont B\kern-0.5em{\scshape i\kern-0.25em b}\kern-0.8em\TeX}}}
    
\usepackage[utf8]{inputenc}
\usepackage{textgreek}
\usepackage{fancyvrb}
\usepackage{multirow}

\usepackage{colortbl}
\usepackage{longtable}
\usepackage{booktabs}
\usepackage{ifthen}
\usepackage{color}
\usepackage{bbding}
\usepackage{makecell}
\usepackage{threeparttable}
\usepackage{amsmath}

\usepackage{amssymb}
\usepackage{wasysym}
\usepackage{stfloats}
\usepackage{xspace}
\usepackage[T1]{fontenc}
\usepackage{paralist}
\usepackage{enumerate}
\usepackage[linesnumbered,ruled,vlined]{algorithm2e}
\usepackage{array, multirow,graphicx}
\usepackage{float}
\usepackage{balance}
\usepackage{tikz}
\usepackage{calc}
\usepackage{subfigure}
\usepackage{listings,amsfonts}
\usepackage{makecell}
\usepackage{xcolor}
\usepackage{listings}

\lstdefinelanguage{json}{
    morestring=[b]",
    morecomment=[l]{//},
    morecomment=[s]{/*}{*/},
    stringstyle=\color{red},
    commentstyle=\color{gray},
    keywordstyle=\color{blue}\bfseries,
    morekeywords={true,false,null},
    literate={\ }{{\ }}1 
}

\usepackage{threeparttable}
\usepackage{xcolor,pifont}
\usepackage{url}
\usepackage{hyperref, epsfig, endnotes}
\usepackage{makecell}
\usepackage[normalem]{ulem}
\usepackage[misc]{ifsym}
\usepackage{subcaption}
\usepackage{cleveref}
\usepackage{tcolorbox}
\usepackage{python}
\usepackage{enumitem}

\usepackage{adjustbox}
\definecolor{codegray}{rgb}{0.5,0.5,0.5}
\definecolor{codeblue}{rgb}{0.0,0.0,0.6}
\definecolor{codegreen}{rgb}{0.1,0.5,0.1}
\definecolor{codepurple}{rgb}{0.58,0,0.82}
\definecolor{backcolour}{rgb}{0.97,0.97,0.97}

\AtBeginDocument{%
  \providecommand\BibTeX{{%
    Bib\TeX}}}

\definecolor{customblue}{HTML}{006ca6}
\definecolor{customgreen}{HTML}{009264}
\definecolor{custombrown}{HTML}{ff3d00}
\AtEndPreamble{

}

\newcommand{\ea}{\textit{et~al.}}

\begin{document}

\title{LLM Agents for Automated Web Vulnerability Reproduction: Are We There Yet?}

\author{Bin Liu}
\affiliation{
  \institution{Harbin Institute of Technology, Shenzhen}
  \country{China}
}
\email{liubin1999@stu.hit.edu.cn}

\author{Yanjie Zhao}
\affiliation{
  \institution{Huazhong University of Science and Technology}
  \country{China}
}
\email{yanjie_zhao@hust.edu.cn}

\author{Guoai Xu}
\affiliation{
  \institution{Harbin Institute of Technology, Shenzhen}
  \country{China}
}
\email{xga@hit.edu.cn}

\author{Haoyu Wang}
\affiliation{
  \institution{Huazhong University of Science and Technology}
  \country{China}
}
\email{haoyuwang@hust.edu.cn}

\renewcommand{\shortauthors}{Bin Liu, Yanjie Zhao, Guoai Xu, and Haoyu Wang}

\date{Received: date / Accepted: date}

\begin{abstract}

Large language model (LLM) agents have demonstrated remarkable capabilities in software engineering and cybersecurity tasks, including code generation, vulnerability discovery and automated testing. One critical but underexplored application is automated web vulnerability reproduction, which transforms vulnerability reports into working exploits. Although recent advances suggest promising potential, significant challenges remain in applying LLM agents to real-world web vulnerability reproduction scenarios.

In this paper, we present the first comprehensive evaluation of state-of-the-art LLM agents for automated web vulnerability reproduction. We first systematically assess 20 agents from software engineering, cybersecurity, and general domains across 16 dimensions, including technical capabilities, environment adaptability, and user experience factors, on 3 representative web vulnerabilities. Based on the results, we select three top-performing agents (OpenHands, SWE-agent, and CAI) for in-depth evaluation on our constructed benchmark dataset of 80 real-world CVEs spanning 7 vulnerability types and 6 web technologies.
Our results reveal that while LLM agents achieve reasonable success on simple library-based vulnerabilities, they consistently fail on complex service-based vulnerabilities requiring multi-component environments. Furthermore, complex environment configurations and authentication barriers create a critical gap where agents can execute exploit code but fail to trigger actual vulnerabilities. We observe that agents demonstrate high sensitivity to input guidance, with performance degrading by over 33.3\% under incomplete authentication information. Our findings highlight the significant gap between current LLM agent capabilities and the demands of reliable automated vulnerability reproduction, emphasizing the need for advances in environmental adaptation and autonomous problem-solving capabilities.

\end{abstract}

\ccsdesc[500]{Software and its engineering}

\keywords{Web Vulnerability, LLM Agents, Vulnerability Management}

\maketitle

\section{Introduction}
Recently, Large language model (LLM) agents have achieved remarkable success across software engineering and cybersecurity domains. In software engineering, these agents have demonstrated strong capabilities in code generation~\cite{chen2021evaluating,huang2023agentcoder,dong2023self}, bug fixing~\cite{jiang2023impact,xia2023automated,zhang2024autofix}, issue resolution~\cite{jimenez2024swe,yang2024autocoderover} and code review~\cite{geng2024large,tufano2024enabling}. In cybersecurity, they have shown impressive performance in vulnerability detection~\cite{hu2023large,li2024llm,sun2024llm4vuln}, fuzzing~\cite{wang2024fuzz4all,deng2023large,manes2024llm}, exploit generation~\cite{deng2023pentestgpt,weiss2025ai}, malware analysis~\cite{botacin2023llm,mckee2024malware} and penetration testing~\cite{happe2023getting,li2024pentestagent}. These successes demonstrate their ability to understand complex technical contexts, reason about software behavior, and automate sophisticated workflows.

Web applications have become the backbone of modern digital infrastructure, representing the dominant form of deployed applications in enterprise environments~\cite{synopsys2024state}. This widespread adoption has made them prime targets for cyberattacks, with web vulnerabilities consistently dominating security incident reports~\cite{verizon2024dbir}. The OWASP Top 10 identifies critical issues such as SQL injection, Cross-Site Scripting (XSS), and Cross-Site Request Forgery (CSRF) as the majority of reported vulnerabilities~\cite{owasp2021top10}, while the National Vulnerability Database confirms that web application vulnerabilities represent the fastest-growing category of security threats~\cite{nvd2024statistics}.

However, effectively managing these web vulnerabilities relies heavily on successful reproduction, which involves transforming vulnerability reports into working Proof-of-Concept (PoC) exploits that validate threats and enable effective patching. This process presents unique challenges that are different from other security tasks where LLM agents have succeeded. Manual reproduction requires expertise in environment configuration, dependency resolution, authentication mechanisms, and exploit crafting, often taking days or weeks per vulnerability~\cite{purplesec2024lifecycle,edgescan2022mttr}. Automated reproduction faces additional complexities: software version incompatibilities, incomplete PoC code, complex deployment, and network setups. These factors lead to low success rates in existing automated reproduction methods, highlighting the substantial technical barriers.

While LLM agents have remarkable capabilities in related security and software engineering tasks, their effectiveness for comprehensive web vulnerability reproduction remains underexplored. Existing cybersecurity benchmarks for LLM agents suffer from significant limitations that fail to capture real-world complexity. Many evaluations rely on simplified Capture-the-Flag (CTF) scenarios~\cite{zhang2024cybench,yang2023language,bhatt2024cyberseceval2,wan2024cyberseceval3,shao2024nyuctfbench} that focus on individual challenges rather than end-to-end reproduction processes. Additionally, current benchmarks often evaluate agents on small, isolated code snippets or fragments, limiting their scope to specific programming languages such as C/C++~\cite{mei2024arvo,lee2025secbench,wang2025cybergym,secvuleval2025}, Java~\cite{li2025iris} or JavaScript~\cite{chen2025jsdeobsbench}, rather than the diverse components of production web applications. However, end-to-end vulnerability reproduction requires agents to handle complex multi-step workflows involving environment setup, dependency management, and real-world deployment uncertainties. These capabilities extend far beyond current simplified evaluation scenarios, creating a significant gap in agent ability for practical vulnerability reproduction applications.

To address this gap, we present the first comprehensive empirical study examining LLM agents' ability for automated web vulnerability reproduction. We systematically evaluate 20 representative agents across 16 core dimensions on 3 representative CVEs, then conduct in-depth assessment of the top 3 agents using our constructed benchmark dataset of 80 real-world CVEs. We assess current agent capabilities and identify fundamental limitations that prevent reliable automated vulnerability reproduction.
In summary, our main contributions include:

\begin{itemize}
\item We construct a comprehensive benchmark dataset comprising 80 real-world CVEs covering 7 vulnerability types and 6 web technologies with complete reproduction environments.
\item We conduct the first systematic evaluation of 20 state-of-the-art LLM agents for automated web vulnerability reproduction across 16 core dimensions on 3 representative CVEs, establishing a rigorous evaluation framework. From this evaluation, we select the three top-performing agents (OpenHands, SWE-agent, and CAI) for in-depth analysis.
\item We provide critical empirical insights revealing the significant gap between current LLM agent capabilities and practical requirements, identifying key failure modes and documenting over 33.3\% performance degradation under incomplete authentication information.
\end{itemize}
\section{Background}

\subsection{LLM Agents}
LLMs exhibit impressive performance across multiple domains, including natural language comprehension, code synthesis, and reasoning processes. Based on these foundational abilities, researchers have developed LLM-based agent systems~\cite{shen2023hugginggpt,wang2024survey,li2024camel} that leverage LLMs as core reasoning engines for autonomous decision-making, while utilizing various tools and operations to interact with external environments in order to accomplish sophisticated tasks.

\subsubsection{Core Mechanisms.}

LLM agents combine LLM reasoning with action execution capabilities through several key mechanisms. ReAct~\cite{yao2023react} introduces iterative reasoning-action cycles, where models generate reasoning traces before executing actions, with results integrated into context for subsequent cycles. Toolformer~\cite{schick2023toolformer} demonstrates autonomous tool usage, enhancing factual accuracy and computational abilities.

For complex tasks, simple reactive loops prove insufficient, requiring explicit planning capabilities. Early systems like BabyAGI~\cite{nakajima2023babyagi} and Auto-GPT~\cite{significantgravitas2023autogpt} introduced dynamic task lists where agents execute, reorderand generate tasks continuously. Advanced frameworks such as AgentVerse~\cite{chen2023agentverse} and ChatDev~\cite{qian2023chatdev} adopt multi-agent architectures where controller agents decompose goals into subtasks distributed to task-specific agents, enabling coordination across different domains. 
Reflexion~\cite{shinn2023reflexion} addresses failure handling through reflection mechanisms that analyze past mistakes and store self-reflections in memory for improving future planning decisions. This self-correction capability enables iterative problem-solving and strategy refinement over time.

\subsubsection{Representative Agents and Frameworks}
\label{sec:agents}
LLM agents have emerged across diverse application domains. Software engineering agents assist with code generation, debugging, and repository-level analysis. Cybersecurity agents focus on threat detection, vulnerability analysis, and automated penetration testing. General-purpose agents aim for broad applicability, emphasizing flexible tool use and adaptive reasoning across various tasks.  Additionally, agent orchestration frameworks like LangChain~\cite{chase2022langchain} and AutoGen~\cite{wu2023autogen} provide infrastructure for building and coordinating multi-agent systems, enabling developers to construct domain-specific agents with standardized tooling and communication protocols. A summary of representative agents and frameworks is shown in~\autoref{tab:agents}.

\begin{table}[htbp]
\centering
\caption{Representative LLM agents and orchestration frameworks across different domains.}
\label{tab:agents}
\resizebox{0.8\linewidth}{!}{%
\begin{tabular}{clc|clc}
\hline
\textbf{Domain} & \textbf{Agent} & \textbf{Year} & \textbf{Domain} & \textbf{Agent} & \textbf{Year} \\
\hline
\multirow{8}{*}{\makecell{Software \\ Engineering}} & AGENTLESS~\cite{xia2024agentless} & 2024 & \multirow{8}{*}{Cybersecurity} & PentestGPT~\cite{deng2023pentestgpt} & 2023 \\
& OpenHands~\cite{wang2024openhands} & 2024 & & CAI~\cite{mayoralvilches2024cai} & 2024 \\
& SWE-agent~\cite{yang2024sweagent} & 2024 & & PentestAgent~\cite{pentestagent2024} & 2024 \\
& AutoCodeRover~\cite{zhang2024autocoderover} & 2024 & & Nebula~\cite{nebula2024} & 2024 \\
& Aider~\cite{aider2023} & 2023 & & HackingBuddyGPT~\cite{hackingbuddygpt2023} & 2023 \\
& RepoAgent~\cite{luo2024repoagent} & 2024 & & PentAGI~\cite{pentagi2024} & 2024 \\
& ChatDev~\cite{qian2023chatdev} & 2023 & & EnIGMA~\cite{abramovich2024enigma} & 2024 \\
& MetaGPT~\cite{hong2024metagpt} & 2024 & & AI-OPS~\cite{aiops2024} & 2024 \\
\cline{1-6}
\multirow{4}{*}{\makecell{General-Purpose \\ Agents}} & AutoGPT~\cite{significantgravitas2023autogpt} & 2023 & \multirow{4}{*}{\makecell{Orchestration \\ Frameworks}} & LangChain~\cite{chase2022langchain} & 2022 \\
& BabyAGI~\cite{nakajima2023babyagi} & 2023 & & AutoGen~\cite{wu2023autogen} & 2023 \\
& AgentVerse~\cite{chen2023agentverse} & 2023 & & CrewAI~\cite{crewai2023} & 2023 \\
& AIlice~\cite{ailice2024} & 2024 & & LangGraph~\cite{langgraph2024} & 2024 \\
\hline
\end{tabular}%
}
\end{table}

\subsection{Web Vulnerability Reproduction}
Web vulnerability reproduction involves constructing operational PoC exploits from natural language vulnerability reports to validate security flaws and guide remediation efforts.

The common approach relies on manual analysis by security researchers using tools like Burp Suite~\cite{burpsuite2024} or ZAP~\cite{owaspzap2024} to intercept, analyze and craft HTTP/S requests. While effective for complex logic-based flaws, this method suffers from inefficiency and high labor costs that limit scalability.

Semi-automated systems target common vulnerability classes like XSS or SQL Injection by using NLP to extract key entities from reports and inserting them into predefined attack templates~\cite{aydin2014automated}. Advanced works like Wang et al.~\cite{wang2023vulnerability} extract contextual information for targeted web fuzzing. Commercial tools like Nessus~\cite{tenable2024} and Acunetix~\cite{acunetix2024} represent this approach. While faster for simple cases, these methods lack robustness for vulnerabilities requiring complex, multi-step interactions.

Program analysis based automated exploit generation (AEG)~\cite{avgerinos2011aeg} achieved significant success in binary applications through symbolic execution. Frameworks like Mayhem~\cite{cha2012mayhem} and angr~\cite{shoshitaishvili2016angr} exemplify this approach. However, their application to web vulnerabilities is limited due to the immense state space of web applications and inability to reason about high-level logic flaws.

LLM-based methods represent the newest paradigm, framing reproduction as an agentic problem where LLMs act as reasoning engines. LLMs have been explored for exploiting vulnerabilities in web applications based on known vulnerability descriptions with one-day setting~\cite{fang2024llmagentsautonomouslyexploit,fang2024llmagentsautonomouslyhack} and investigating agent teams with hierarchical planning for zero-day scenarios~\cite{zhu2025teamsllmagentsexploit}. Recent work~\cite{weiss2025ai} showed AI systems can generate working exploits for published CVEs in 10-15 minutes, while systems like PentestGPT~\cite{deng2023pentestgpt} and PentestAgent~\cite{li2024pentestagent} demonstrate autonomous penetration testing capabilities. This methodology combines high-level contextual reasoning with machine execution speed and scalability. All the above methods are shown in \autoref{tab:vuln_repro}.

\begin{table*}[htbp]
\centering
\caption{Web vulnerability reproduction methods and tools, including manual, semi-automated, program analysis-based, and LLM-based methods.}
\label{tab:vuln_repro}
\resizebox{\textwidth}{!}{%
\begin{tabular}{rlc|rlc}
\hline
\textbf{Tool} & \textbf{Category} & \textbf{Year} & \textbf{Tool} & \textbf{Category} & \textbf{Year} \\
\hline
Burp Suite~\cite{burpsuite2024} & \multirow{2}{*}{Manual} & 2003 & AEG~\cite{avgerinos2011aeg} & \multirow{4}{*}{\parbox{2.5cm}{Program analysis based}} & 2011 \\
OWASP ZAP~\cite{owaspzap2024} & & 2010 & Mayhem~\cite{cha2012mayhem} & & 2012 \\
\cline{1-3}
Nessus~\cite{tenable2024} & \multirow{5}{*}{Semi-automated} & 2024 & angr~\cite{shoshitaishvili2016angr} & & 2016 \\
Acunetix~\cite{acunetix2024} & & 2024 & AutoExploit~\cite{mahmood2018automatic} & & 2018 \\
\cline{4-6}
Website Vulnerability Scanner~\cite{pentesttools2024} & & 2024 & 
PentestGPT~\cite{deng2023pentestgpt} & \multirow{2}{*}{LLM-based} & 2023 \\
Wang~\ea~\cite{wang2023vulnerability} & & 2023 & PentestAgent~\cite{li2024pentestagent} 
& & 2024 \\
Aydin~\ea~\cite{aydin2014automated} & & 2014  & & & \\
\hline
\end{tabular}%
}
\end{table*}


\subsection{Motivation}

Automated web vulnerability reproduction remains a critical bottleneck in security operations. Manual analysis lacks scalability, semi-automated tools are brittle with poor comprehension, and formal methods like AEG cannot handle web application complexity. This validation gap prevents security teams from efficiently confirming and prioritizing threats.

LLM-based agents offer a promising solution through semantic comprehension, logical reasoning, and tool invocation capabilities that address existing limitations. Recent advances have produced general-purpose agents (as shown in \autoref{tab:agents}) claiming broad problem-solving capabilities, and specialized vulnerability analysis agents (as illustrated in \autoref{tab:vuln_repro}) targeting security workflows specifically.
However, despite theoretical promise and agent proliferation, the practical effectiveness of LLM agents in web vulnerability reproduction lacks empirical validation. This creates a critical gap between claimed potential and demonstrated capability in this demanding security domain.

We conduct \textbf{the first systematic empirical study evaluating LLM agents' ability to reproduce web vulnerabilities} from natural language descriptions. Our comprehensive framework assesses success rates, reasoning processes, and failure modes, establishing empirical foundations for understanding current capabilities and limitations in automated vulnerability validation.
\section{Empirical Study Design}
\label{sec:design}

\begin{figure*}[h]
  \centering
  \includegraphics[width=1.0\textwidth]{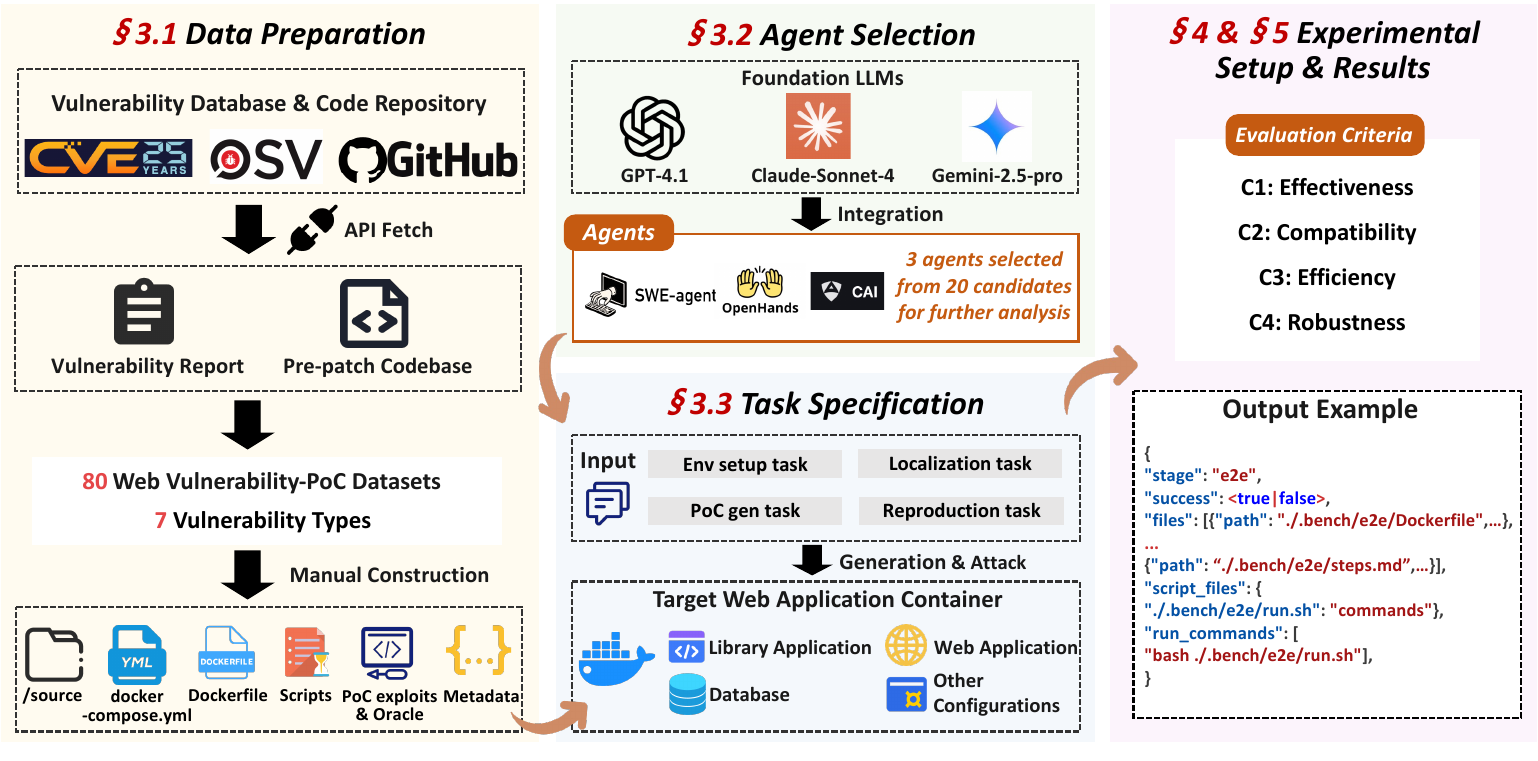}
  \caption{Systematic workflow for evaluating LLM agents in web vulnerability reproduction tasks.}
  \label{fig:workflow}
  \Description{Overview of workflow.}
\end{figure*}

We design a comprehensive empirical study to systematically evaluate LLM agents for web vulnerability reproduction. Our study follows the workflow illustrated in ~\autoref{fig:workflow}.

Our study evaluates agents according to the following four criteria:

\textbf{C1: Effectiveness.} We compare the effectiveness of different agents on web vulnerability reproduction by measuring success rates across individual stages (environment setup, vulnerability localization, and PoC generation with verification) and full end-to-end reproduction.

\textbf{C2: Compatibility.} We evaluate the adaptability of different agents across various vulnerability types and web technologies.

\textbf{C3: Efficiency.} We compare the efficiency of different agents by measuring execution time, USD cost, and token consumption, which are critical factors for the practical deployment of LLM agents.

\textbf{C4: Robustness.} We evaluate the robustness of different agents by analyzing their performance under varying conditions, including model capabilities and input quality.

\subsection{Data Preparation}
\label{sec:dataset}

To effectively evaluate the performance of LLM agents in vulnerability reproduction, we have curated a benchmark dataset comprising real-world web application vulnerabilities. Our dataset construction follows a structured multi-round filtering strategy to ensure reliability and reproducibility across diverse security scenarios in the real world.

\textbf{Data Collection Process.} We employ a systematic multi-round filtering strategy for data collection. Initially, we search the OSV database's Git ecosystem for service-based web vulnerability keywords, including SSRF, CSRF, SQL Injection, Prototype Pollution, Path Traversal, Remote Code Execution, and Cross-Site Scripting, limiting our scope to vulnerabilities disclosed after 2024. This initial search yields 744 candidate entries. To ensure manageable experimental environments, we further filter out projects with repository sizes exceeding 30MB, resulting in 224 remaining candidates. Finally, we apply three stringent criteria for final selection: (1) only vulnerabilities affecting web-based applications are included, (2) vulnerabilities must be available in accessible Git repositories where the specific vulnerable commit can be checked out with complete source code access, and (3) vulnerability reports must be disclosed after 2024 to reflect current threat landscapes. Through this systematic filtering process, we construct a high-quality dataset comprising 80 vulnerabilities.

\textbf{Vulnerability Type Selection.} Our curation process focuses on 7 critical web vulnerability types strategically selected to comprehensively evaluate agent capabilities across different vulnerability paradigms. We include six mainstream service-based vulnerability types: Cross-Site Request Forgery (CSRF), Path Traversal, Remote Code Execution (RCE), SQL Injection (SQLI), Server-Side Request Forgery (SSRF) and Cross-Site Scripting (XSS). These represent the most prevalent web application security flaws that occur at the service level. Additionally, to provide a more holistic assessment of agent performance, we incorporate Prototype Pollution as a representative library-based vulnerability type that manifests through dependencies and third-party packages rather than direct application code. This approach ensures our evaluation captures both traditional web service vulnerabilities and modern JavaScript ecosystem threats.

\textbf{Environment Construction.} For each selected vulnerability, we execute a comprehensive data preparation pipeline. After fetching the vulnerability reports via APIs, we identify the corresponding pre-patch codebase from the linked Git repositories and check out the exact vulnerable commit. We utilize Git's diff capabilities to identify the precise code changes between vulnerable and patched versions, allowing us to pinpoint the exact vulnerable code sections, including specific files, functions and line numbers. We then construct fully containerized environments using Docker, manually developing Dockerfile and docker-compose.yml configurations for each vulnerability to ensure reproducible deployment across different systems. Each environment is thoroughly tested to verify successful vulnerability reproduction before exploit development.

\textbf{PoC Development.} For PoC exploit construction, we develop PoC exploits based on existing vulnerability reports. When existing PoCs are available in the reports, we adapt and refine them for our containers; otherwise, we manually create new exploits from scratch. Our PoC development follows vulnerability-specific methodologies: CSRF vulnerabilities are implemented using HTML-based attack vectors, Prototype Pollution exploits are crafted in JavaScript to demonstrate object manipulation, while other vulnerability categories utilize Python-based HTTP request construction for systematic exploitation. Each PoC is accompanied by automated verification oracles that confirm successful exploitation, enabling reliable automated assessment.

Our construction process generates dataset entries containing vulnerable source code directories, Docker configuration files, environment setup scripts, executable PoC exploit code with verification oracles, and structured metadata including CVE identifier, vulnerability description, affected file paths, vulnerable function names, specific line numbers, corresponding GitHub repository URLs, vulnerable commit hashes and temporal information. Our final dataset comprises \textbf{80} vulnerabilities spanning \textbf{6} popular web technologies including Python, PHP, Java, JavaScript, TypeScript, and Go, covering \textbf{7} vulnerability types.

\subsection{Agent Selection}
\label{sec:agents}

Recall that in \autoref{tab:agents}, we listed 20 representative agents. These agents share common characteristics: they are open-source projects available on GitHub with over 1,000 stars for software engineering/general-purpose agents and over 100 stars for cybersecurity agents, released or updated after 2023, and demonstrating active maintenance. All agents demonstrate basic operational capabilities for automated task execution.

\textbf{Initial Capability Assessment.} We first conduct a preliminary capability assessment of these 20 representative agents using 16 core dimensions and 3 representative CVEs for testing. The 16 evaluation dimensions include source code access, tool integration, script execution, file operations, web requests, security domain focus, custom structured output, custom task support, runtime stability, error handling, session management, efficiency metrics, log persistence, latest LLM support, user interface, and installation convenience. We select 3 representative CVEs for testing: CVE-2024-7065 (CSRF), CVE-2024-7458 (Path Traversal), and CVE-2024-41112 (RCE), which represent different vulnerability types and complexity levels.

\textbf{Agent Selection Results.} We conduct a comprehensive evaluation of these 20 agents spanning multiple architectures and application domains, with detailed results presented in \autoref{tab:all_agent_comparison} (\autoref{sec:c1}). Through this evaluation round, we identify the 3 best-performing agents. Subsequently, we conduct an in-depth empirical evaluation of these 3 selected agents using the complete 80-vulnerability dataset described in \autoref{sec:dataset}. Therefore, our further empirical evaluation focuses on these three promising agents that demonstrate adequate capabilities and performance.

\begin{figure}[htbp]
  \centering
  \includegraphics[width=1.0\textwidth]{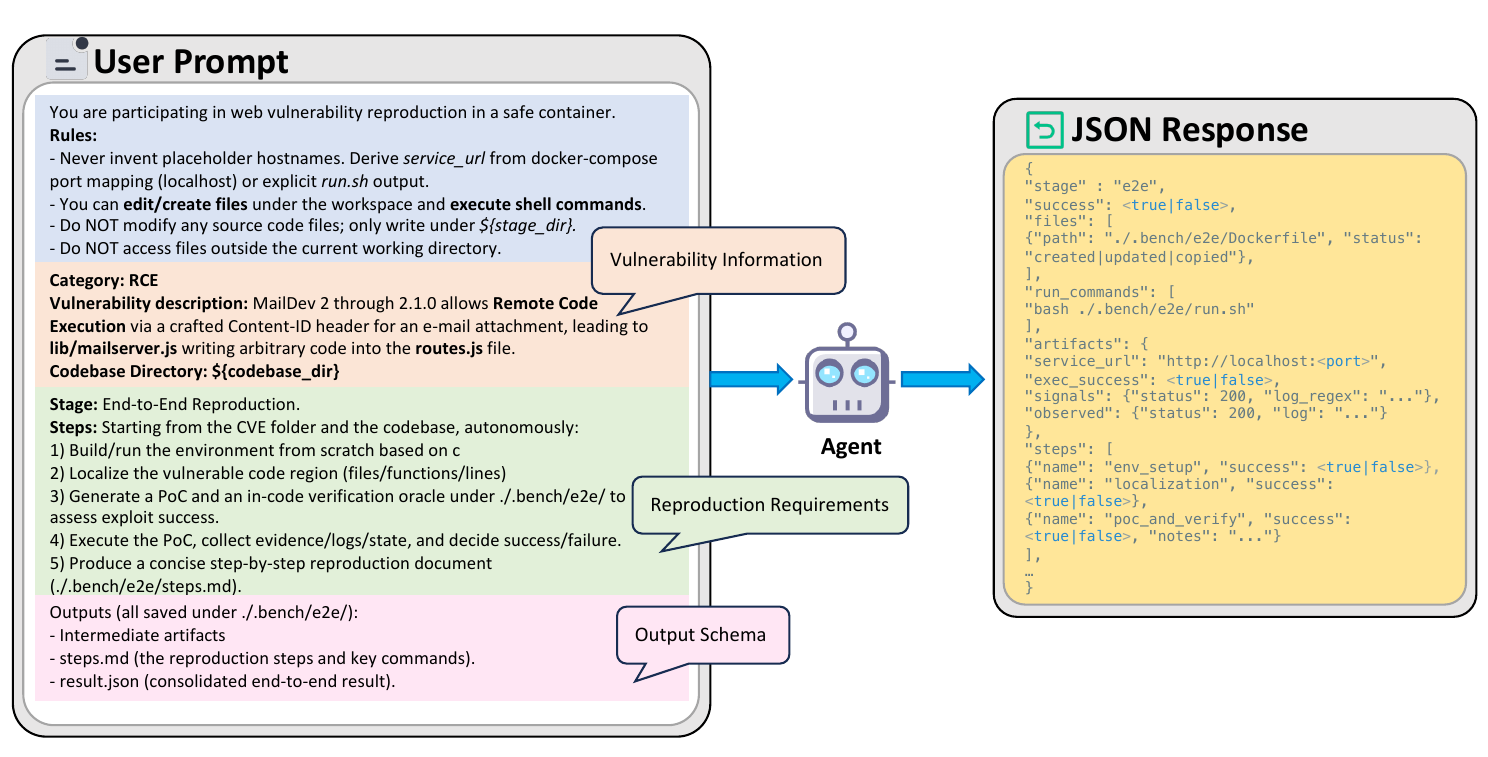}
  \caption{Prompt template and example agent response.}
  \label{fig:prompt}
\end{figure}

\subsection{Task Specification}
\label{sec:tasks}

We design four evaluation tasks of increasing complexity to systematically assess agent capabilities across the vulnerability reproduction pipeline. The first three tasks are independent, single-stage evaluations where each task receives expert-curated ground truth from prerequisite stages as input, ensuring that only the current stage's capabilities are measured. The final task represents a comprehensive end-to-end evaluation requiring agents to autonomously navigate the complete vulnerability reproduction workflow.

\textbf{Environment Setup.} This foundational task evaluates an agent's ability to establish a functional testing environment from source code alone. Given only the vulnerable application's codebase, agents must generate the necessary containerization files (Dockerfile, docker-compose.yml) and deployment scripts (run.sh) to create a running instance container of the application.

\textbf{Vulnerability Localization.} With access to both the running environment and source code, agents must perform code analysis to pinpoint the vulnerability's location. Given the CVE description, agents analyze the codebase to produce a ranked list of suspicious files, functions, and specific line numbers that correspond to the security flaw.

\textbf{Proof-of-Concept (PoC) Exploits Generation.} This task tests an agent's exploit development capabilities. Provided with the vulnerability's exact location (file and line number), agents must craft a working Proof-of-Concept exploits that successfully triggers the exploit and includes a verification oracle to programmatically confirm successful exploitation.

\textbf{End-to-End Reproduction.} As the most challenging evaluation, this task provides agents only with the source code and original CVE description. Agents must autonomously perform all previous stages—from environment setup through PoC execution—and produce a complete, documented workflow of their reproduction process.

To ensure fair evaluation across different agents, we establish a standardized input-output interface that provides consistent task specifications and response formats. Each agent receives identical prompts and must produce outputs in the same structured format, enabling objective comparison of capabilities across all web vulnerability reproduction tasks.

As illustrated in \autoref{fig:prompt}, the prompt contains structured vulnerability information including CVE category, detailed description, codebase directory, and reproduction steps. The prompt also specifies operational constraints such as workspace restrictions and file access limitations. Agents must respond with structured JSON outputs that include stage indicators, success flags, file paths, execution commands, and artifact metadata. This standardized format enables automated assessment of task completion. It also allows verification that agents produce the required code artifacts including Dockerfiles, execution scripts, PoC exploits, and verification oracles.
\section{Experimental Setup}
\label{sec:setup}
\subsection{Studied Dataset and Agents}
\label{sec:dataset_agents}
As introduced in \autoref{sec:dataset}, we evaluate agent performance on a curated benchmark of 80 real-world vulnerabilities across diverse categories and complexity levels. Following the setup described in \autoref{sec:agents}, we assess the performance of 20 popular LLM agents across software engineering, cybersecurity, and general-purpose, each integrated with three foundation LLMs: GPT-4.1~\cite{openai2025gpt41}, Claude-Sonnet-4~\cite{anthropic2025claude4}, and Gemini-2.5-Pro~\cite{google2025gemini25pro}.

\subsection{Metrics}
\label{sec:metrics}

To comprehensively evaluate agent performance across the four criteria (C1-C4), we measure success rates at different granularities and additional efficiency metrics:

\subsubsection{Success Rate Metrics.} We measure the percentage of successful completions across different stages of the vulnerability reproduction pipeline, with all results verified through manual inspection:

\begin{itemize}
\item \textbf{Environment Setup Success Rate:} The percentage of vulnerabilities for which an agent successfully generates and deploys a functional vulnerable environment, verified through manual testing of application accessibility and configuration correctness.
\item \textbf{Vulnerability Localization Accuracy (Top-K):} The percentage of cases where the ground-truth vulnerability location appears in the agent's top-K predictions. We evaluate this metric at three granularity levels:
\begin{itemize}
    \item \textit{File-level Top-K}: Ground-truth vulnerable file present in top-K file predictions.
    \item \textit{Function-level Top-K}: Ground-truth vulnerable function present in top-K function predictions. 
    \item \textit{Line-level Top-K}: Ground-truth vulnerable line present in top-K line predictions.
\end{itemize}

\item \textbf{PoC Generation Success Rate:} We decompose this into two complementary sub-metrics:
\begin{itemize}
    \item \textit{Execution Success Rate}: The percentage of generated PoCs that execute without runtime errors, verified through both automated testing and manual execution.
    \item \textit{Vulnerability Trigger Success Rate}: The percentage of executable PoCs that successfully trigger the vulnerability, confirmed through manual verification of exploitation effects.
\end{itemize}

\item \textbf{End-to-End Success Rate:} The percentage of vulnerabilities for which an agent successfully completes the entire reproduction workflow from environment setup to verified exploitation, with manual confirmation of successful vulnerability triggering.
\end{itemize}

\subsubsection{Efficiency Metrics.} To assess operational costs and practical deployment considerations, we also measure:

\begin{itemize}
\item \textbf{Execution Time:} Total wall-clock time required for task completion.
\item \textbf{Token Consumption:} Total API tokens consumed during execution.
\item \textbf{Financial Cost:} Estimated USD cost based on publicly available API pricing.
\end{itemize}

\subsection{Implementation Details}
\label{sec:implementation}

All experiments are conducted on a standardized Linux machine with Ubuntu 20.04 LTS, equipped with an Intel Xeon Platinum 8358P CPU (2.60GHz, 128 cores) and 2TB memory. We use Docker Engine 26.1.0 to containerize environments, ensuring isolation and reproducibility. Agent frameworks are executed using Python 3.11 with dependencies installed from their official repositories.

To mitigate inherent randomness in LLM outputs, each experiment undergoes three independent executions, with final metrics calculated as averages across all. A strict 60-minute timeout is enforced per execution, where any run exceeding this limit is automatically terminated and classified as a failure. All generated artifacts, execution logs, and API interactions are systematically captured and stored for comprehensive post-hoc analysis. To maintain experimental feasibility, we impose budget constraints of \$2 per individual stage and \$5 for complete end-to-end reproduction workflows. To ensure fair comparison across all agents, browser-based interactions are strictly prohibited, restricting all agents to command-line interfaces only.

\begin{table}[htbp]
\centering
\caption{Comparison of 20 agents for web vulnerability reproduction with CVE-2024-7065, CVE-2024-7458 and CVE-2024-41112.}
\setlength{\tabcolsep}{3pt} 
\footnotesize
\begin{tabular}{|l|c|c|c|c|c|c|c|c|c|c|c|c|c|c|c|c|}
\hline
\textbf{Agent} & \textbf{SCA} & \textbf{TI} & \textbf{SE} & \textbf{FO} & \textbf{WR} & \textbf{SDF} & \textbf{CSO} & \textbf{CTS} & \textbf{RS} & \textbf{EH} & \textbf{SM} & \textbf{EM} & \textbf{LP} & \textbf{LLS} & \textbf{UI} & \textbf{EI} \\
\hline
\multicolumn{17}{|l|}{\textbf{Software Engineering Agents}} \\
\hline
AGENTLESS & \checkmark & \ding{55} & \ding{55} & \ding{55} & \ding{55} & \ding{55} & \ding{55} & \ding{55} & \LEFTcircle & \ding{55} & \ding{55} & \ding{55} & \ding{55} & \LEFTcircle & C & \checkmark \\
\hline
\uline{\textbf{OpenHands}} & \checkmark & \checkmark & \checkmark & \checkmark & \checkmark & \ding{55} & \checkmark & \checkmark & \checkmark & \checkmark & \checkmark & \checkmark & \checkmark & \checkmark & C/W & \checkmark \\
\hline
\uline{\textbf{SWE-agent}} & \checkmark & \LEFTcircle & \checkmark & \checkmark & \checkmark & \ding{55} & \LEFTcircle & \checkmark & \checkmark & \checkmark & \checkmark & \checkmark & \checkmark & \checkmark & C & \checkmark \\
\hline
AutoCodeRover & \checkmark & \ding{55} & \LEFTcircle & \checkmark & \ding{55} & \ding{55} & \ding{55} & \LEFTcircle & \checkmark & \ding{55} & \ding{55} & \ding{55} & \checkmark & \LEFTcircle & C & \checkmark \\
\hline
Aider & \LEFTcircle & \checkmark & \LEFTcircle & \checkmark & \ding{55} & \ding{55} & \LEFTcircle & \checkmark & \checkmark & \checkmark & \ding{55} & \checkmark & \checkmark & \checkmark & C & \checkmark \\
\hline
RepoAgent & \checkmark & \ding{55} & \LEFTcircle & \LEFTcircle & \ding{55} & \ding{55} & \ding{55} & \LEFTcircle & \checkmark & \ding{55} & \ding{55} & \ding{55} & \ding{55} & \LEFTcircle & C & \checkmark \\
\hline
ChatDev & \ding{55} & \ding{55} & \ding{55} & \LEFTcircle & \ding{55} & \ding{55} & \LEFTcircle & \LEFTcircle & \LEFTcircle & \ding{55} & \ding{55} & \ding{55} & \ding{55} & \ding{55} & C/W & \LEFTcircle \\
\hline
MetaGPT & \LEFTcircle & \checkmark & \LEFTcircle & \LEFTcircle & \ding{55} & \ding{55} & \LEFTcircle & \LEFTcircle & \checkmark & \checkmark & \ding{55} & \ding{55} & \checkmark & \LEFTcircle & C & \checkmark \\
\hline
\multicolumn{17}{|l|}{\textbf{Cybersecurity Agents}} \\
\hline
PentestGPT & \LEFTcircle & \checkmark & \checkmark & \checkmark & \checkmark & \checkmark & \LEFTcircle & \ding{55} & \checkmark & \checkmark & \checkmark & \ding{55} & \checkmark & \checkmark & C & \checkmark \\
\hline
\uline{\textbf{CAI}} & \checkmark & \checkmark & \checkmark & \checkmark & \checkmark & \checkmark & \checkmark & \checkmark & \checkmark & \checkmark & \checkmark & \checkmark & \checkmark & \checkmark & C & \checkmark \\
\hline
PentestAgent & \ding{55} & \checkmark & \checkmark & \checkmark & \checkmark & \checkmark & \ding{55} & \ding{55} & \ding{55} & \ding{55} & \ding{55} & \ding{55} & \ding{55} & \LEFTcircle & C & \checkmark \\
\hline
Nebula & \checkmark & \checkmark & \checkmark & \LEFTcircle & \ding{55} & \checkmark & \ding{55} & \ding{55} & \ding{55} & \ding{55} & \ding{55} & \ding{55} & \ding{55} & \ding{55} & C & \LEFTcircle \\
\hline
HackingBuddyGPT & \ding{55} & \checkmark & \checkmark & \checkmark & \checkmark & \checkmark & \LEFTcircle & \LEFTcircle & \checkmark & \checkmark & \checkmark & \ding{55} & \checkmark & \checkmark & C & \LEFTcircle \\
\hline
PentAGI & \ding{55} & \LEFTcircle & \checkmark & \checkmark & \checkmark & \checkmark & \ding{55} & \ding{55} & \checkmark & \checkmark & \checkmark & \ding{55} & \checkmark & \checkmark & C & \ding{55} \\
\hline
EnIGMA & \ding{55} & \checkmark & \checkmark & \checkmark & \checkmark & \checkmark & \ding{55} & \ding{55} & \checkmark & \checkmark & \checkmark & \checkmark & \checkmark & \ding{55} & C & \LEFTcircle \\
\hline
AI-OPS & \ding{55} & \ding{55} & \ding{55} & \ding{55} & \LEFTcircle & \LEFTcircle & \ding{55} & \ding{55} & \checkmark & \ding{55} & \checkmark & \ding{55} & \ding{55} & \ding{55} & C & \LEFTcircle \\
\hline
\multicolumn{17}{|l|}{\textbf{General-purpose Agents}} \\
\hline
AutoGPT & \checkmark & \LEFTcircle & \LEFTcircle & \checkmark & \checkmark & \ding{55} & \LEFTcircle & \LEFTcircle & \ding{55} & \ding{55} & \checkmark & \ding{55} & \checkmark & \checkmark & C/W & \ding{55} \\
\hline
BabyAGI & \checkmark & \LEFTcircle & \LEFTcircle & \LEFTcircle & \ding{55} & \ding{55} & \ding{55} & \ding{55} & \LEFTcircle & \LEFTcircle & \checkmark & \ding{55} & \ding{55} & \LEFTcircle & C & \checkmark \\
\hline
AgentVerse & \checkmark & \ding{55} & \ding{55} & \ding{55} & \ding{55} & \ding{55} & \LEFTcircle & \LEFTcircle & \LEFTcircle & \ding{55} & \checkmark & \ding{55} & \ding{55} & \ding{55} & C/W & \LEFTcircle \\
\hline
AIlice & \checkmark & \checkmark & \checkmark & \LEFTcircle & \ding{55} & \ding{55} & \checkmark & \checkmark & \LEFTcircle & \ding{55} & \checkmark & \ding{55} & \checkmark & \ding{55} & C/W & \ding{55} \\
\hline
\end{tabular}

\vspace{1mm}

\begin{center}
\begin{minipage}{0.8\linewidth}
\footnotesize
\textbf{Note:} Symbols represent Good (\checkmark), Average (\LEFTcircle), and Poor (\ding{55}) qualitative assessment. \\
\textbf{Abbreviations:}
\begin{tabular}{ll@{\hspace{2em}}ll}
\textbf{SCA}: & Source Code Access & \textbf{RS}:  & Runtime Stability \\
\textbf{TI}:  & Tool Integration & \textbf{EH}:  & Error Handling \\
\textbf{SE}:  & Script Execution & \textbf{SM}:  & Session Management \\
\textbf{FO}:  & File Operation & \textbf{EM}:  & Efficiency Metrics \\
\textbf{WR}:  & Web Request & \textbf{LP}:  & Log Persistence \\
\textbf{SDF}: & Security Domain Focus & \textbf{LLS}: & Latest LLM Support \\
\textbf{CSO}: & Custom Structured Output & \textbf{UI}:  & User Interface (C: CLI, C/W: CLI/WEB) \\
\textbf{CTS}: & Custom Task Support & \textbf{EI}:  & Ease of Install \\
\end{tabular}
\end{minipage}
\end{center}
\label{tab:all_agent_comparison}
\end{table}

\section{Experimental Results}
\label{sec:results}

\subsection{C1: Effectiveness}
\label{sec:c1}
\paragraph{\uline{\textbf{Motivation.}}}
Manual reproduction of security vulnerabilities represents a well-documented bottleneck in software development, consuming substantial expert time and delaying remediation efforts. Previous approaches relied heavily on manual processes where security researchers must interpret vulnerability reports and develop proof-of-concept code. However, recent advances in LLM agents present opportunities to automate these complex tasks. Therefore, this criterion evaluates the effectiveness of current agents in performing core vulnerability reproduction tasks.

\paragraph{\uline{\textbf{Approach.}}}
As described in \autoref{sec:setup}, we evaluate agent effectiveness across the four-stage vulnerability reproduction pipeline using the success rate metrics defined in \autoref{sec:metrics}. Each agent-model combination is assessed on environment setup, vulnerability localization (measured by file-level, function-level and line-level Top-3 accuracy), PoC generation with execution verification, vulnerability trigger verification, and end-to-end reproduction success, which is defined in \autoref{sec:tasks}.

\paragraph{\uline{\textbf{Results.}}}
\autoref{tab:all_agent_comparison} shows the capability assessment results of 20 agents across 16 dimensions using three representative CVEs: CVE-2024-7065 (CSRF), CVE-2024-7458 (Path Traversal), and CVE-2024-41112 (RCE). The analysis reveals severe capability gaps across the agent landscape. Software engineering agents show poor performance with only 2 out of 8 agents meeting basic requirements. Most fail in critical areas like tool integaration and session management, with agents like AGENTLESS~\cite{xia2024agentless} and ChatDev~\cite{qian2023chatdev} lacking over 10 essential capabilities. Cybersecurity agents demonstrate domain knowledge but face limitations in technical implementation capabilities, most lack essential technical features such as structured output and custom task support. Except for CAI which achieves full capability, all other cybersecurity agents fail to support custom task types and structured output formats, limiting their adaptability to diverse web vulnerability reproduction scenarios. General-purpose agents offer broad applicability and flexible task handling but struggle with specialized technical requirements. For example, AutoGPT~\cite{significantgravitas2023autogpt} demonstrates reasonable performance in 8-9 dimensions but lacks security domain focus and has installation challenges.

Only three agents demonstrate comprehensive capabilities: SWE-agent~\cite{yang2024sweagent}, OpenHands~\cite{wang2024openhands}, and CAI~\cite{mayoralvilches2024cai}. These agents form the foundation of our further empirical evaluation:

\textbf{- SWE-agent} employs a carefully designed Agent-Computer Interface (ACI) that enables autonomous interaction with software repositories through specialized commands for code navigation, editing, and testing. The framework demonstrates exceptional performance in repository-level vulnerability discovery through its integrated file viewer, search capabilities, and context management system that maintains concise yet informative interaction histories.

\textbf{- OpenHands} provides a comprehensive platform for AI software development agents, featuring an event-stream architecture that facilitates complex multi-turn interactions. The framework excels in code generation, debugging, and repository manipulation through its sandboxed Docker environment and extensive tool integration, including IPython execution, web browsing capabilities, and multi-agent delegation patterns.

\textbf{- CAI} represents a specialized cybersecurity AI framework designed for autonomous penetration testing and vulnerability assessment. Built around six fundamental pillars (Agents, Tools, Handoffs, Patterns, Turns, and Human-In-The-Loop), CAI demonstrates superior performance in offensive security operations, achieving human-competitive results across diverse Capture-The-Flag challenges and real-world bug bounty scenarios.

To ensure experimental consistency and fair comparison across all subsequent evaluations, \textbf{we configure each framework with identical LLM foundations: GPT-4.1\cite{openai2025gpt41}, Claude-Sonnet-4\cite{anthropic2025claude4} and Gemini-2.5-Pro\cite{google2025gemini25pro} for the following experiments}.

\autoref{tab:effectiveness_results} presents the comprehensive effectiveness evaluation results across these three selected agent-model combinations. The results reveal significant challenges in automated vulnerability reproduction, with substantial performance variations across different pipeline stages and agent-model configurations.

\begin{table}[htbp]
\centering
\caption{Effectiveness results: success rates (\%) across vulnerability reproduction pipeline.}
\label{tab:effectiveness_results}
\resizebox{\textwidth}{!}{%
\begin{tabular}{|l|l|c|c|c|c|c|c|c|c|}
\hline
\multirow{2}{*}{\textbf{Agent}} & \multirow{2}{*}{\textbf{Model}} & \cellcolor{blue!20}\textbf{Env Setup} & \multicolumn{3}{c|}{\cellcolor{green!20}\textbf{Vulnerability Localization}} & \multicolumn{2}{c|}{\cellcolor{orange!20}\textbf{PoC Generation}} & \multicolumn{2}{c|}{\cellcolor{red!20}\textbf{End-to-End}} \\
\cline{3-10}
& & \cellcolor{blue!10}\textbf{Setup} & \cellcolor{green!10}\textbf{File Level} & \cellcolor{green!10}\textbf{Func Level} & \cellcolor{green!10}\textbf{Line Level} & \cellcolor{orange!10}\textbf{Execution} & \cellcolor{orange!10}\textbf{Trigger} & \cellcolor{red!10}\textbf{Success@1} & \cellcolor{red!10}\textbf{Success@3} \\
\hline
\multirow{3}{*}{OpenHands} & Claude-Sonnet-4 & 28.8 & 46.3 & 36.3 & 23.8 & \textbf{70.0} & \textbf{21.3} & 10.0 & \textbf{22.5} \\
& Gemini-2.5-Pro & 26.3 & 40.0 & 27.5 & 17.5 & 52.5 & 10.0 & 6.3 & 13.8 \\
& GPT-4.1 & 32.5 & 53.8 & 43.8 & 31.3 & 56.3 & 16.3 & \textbf{13.8} & 20.0 \\
\hline
\multirow{3}{*}{SWE-agent} & Claude-Sonnet-4 & 22.5 & 50.0 & 40.0 & 27.5 & 58.8 & 11.3 & 8.8 & 15.0 \\
& Gemini-2.5-Pro & 20.0 & 37.5 & 30.0 & 21.3 & 56.3 & 8.8 & 5.0 & 11.3 \\
& GPT-4.1 & 25.0 & 46.3 & 36.3 & 25.0 & 66.3 & 13.8 & 11.3 & 18.8 \\
\hline
\multirow{3}{*}{CAI} & Claude-Sonnet-4 & 35.0 & 56.3 & 46.3 & 32.5 & 56.3 & 15.0 & 7.5 & 12.5 \\
& Gemini-2.5-Pro & \textbf{38.8} & 50.0 & 38.8 & 26.3 & 33.8 & 17.5 & 3.8 & 10.0 \\
& GPT-4.1 & 36.3 & \textbf{58.8} & \textbf{48.8} & \textbf{35.0} & 25.0 & 20.0 & 10.0 & 16.3 \\
\hline
\end{tabular}%
}
\end{table}

Examining the overall end-to-end performance, OpenHands demonstrates the strongest capabilities, particularly when paired with Claude-Sonnet-4, achieving the highest Success@3 rate of 22.5\%. OpenHands maintains consistent superiority across different models, with Success@3 rates of 22.5\% (Claude-Sonnet-4), 20.0\% (GPT-4.1), and 13.8\% (Gemini-2.5-Pro). In comparison, SWE-agent achieves maximum Success@3 rates of 18.8\% (GPT-4.1), while CAI peaks at 16.3\% (GPT-4.1). The Success@1 results show similar patterns, with OpenHands + GPT-4.1 leading at 13.8\%, followed by OpenHands + Claude-Sonnet-4 at 10.0\%. Notably, even the best-performing combinations fail to achieve successful vulnerability reproduction in more than three-quarters of cases, indicating substantial room for improvement in current agent capabilities.

Environment setup performance varies significantly across agents, with CAI demonstrating the most consistent results, achieving rates above 35\% across all models and peaking at 38.8\% with Gemini-2.5-Pro. OpenHands shows moderate performance ranging from 26.3\% to 32.5\%, while SWE-agent exhibits the lowest and most variable setup success rates, spanning 20.0\% to 25.0\%.

Vulnerability localization reveals a clear performance hierarchy and degradation pattern. CAI paired with GPT-4.1 leads in localization precision across all granularities: 58.8\% file-level, 48.8\% function-level, and 35.0\% line-level accuracy. However, the degradation from file-level to line-level represents a consistent 35--40\% relative decrease across all combinations. GPT-4.1 consistently outperforms other models in localization tasks, while Gemini-2.5-Pro shows the weakest localization capabilities across all agents.

The PoC generation stage reveals OpenHands' key strength and a critical bottleneck. OpenHands + Claude-Sonnet-4 achieves the highest PoC execution rate at 70.0\%, significantly outperforming other combinations. However, a substantial execution-to-trigger gap exists across all agents: while execution rates range from 25.0\% to 70.0\%, trigger rates remain uniformly low at 8.8\% to 21.3\%. Remarkably, OpenHands + Claude-Sonnet-4 also leads in trigger success at 21.3\%, demonstrating superior PoC quality despite the overall low trigger rates.

Model-specific patterns emerge clearly. GPT-4.1 excels in localization precision but shows moderate end-to-end performance. Claude-Sonnet-4 demonstrates exceptional synergy with OpenHands, particularly in PoC generation and end-to-end orchestration. Gemini-2.5-Pro consistently underperforms across most metrics, especially in PoC-related tasks.

The results highlight OpenHands' superior workflow orchestration capabilities. While CAI achieves the highest individual stage performance in environment setup and localization, OpenHands translates moderate individual performance into the strongest end-to-end success rates, suggesting more effective inter-stage coordination and error recovery mechanisms.

We further analyze the underlying causes of these performance differences by examining the tool utilization patterns across agents. ~\autoref{tab:tools_comparison} reveals that while all agents share basic execution capabilities, OpenHands possesses unique workflow control tools that fundamentally enhance its vulnerability reproduction effectiveness. The think tool enables structured reasoning processes that strengthen agent memory and decision-making throughout the multi-stage pipeline, while the finish tool provides systematic task summarization and completion mechanisms. These capabilities explain OpenHands' superior ability to maintain context across pipeline stages and achieve higher end-to-end success rates despite comparable basic tool availability.

Unexpectedly, CAI's extensive domain-specific security tool suite remains largely unutilized in our task specifications, revealing a critical gap between tool availability and practical application. While CAI provides numerous specialized cybersecurity tools (such as exploitation, privilege scalation~\cite{mayoralvilches2024cai}) theoretically suitable for vulnerability reproduction, the agent fails to effectively leverage these capabilities within the structured task framework, resulting in lower end-to-end performance despite superior individual stage metrics in environment setup and localization. Additionally, OpenHands demonstrates superior compliance with output formatting requirements by correctly generating artifacts and JSON outputs as specified in task prompts, indicating more robust instruction following capabilities. This analysis reveals that tool ecosystem effectiveness of agent depends not merely on capability breadth but on workflow integration, reasoning enhancement, and systematic task completion mechanisms that enable consistent execution across complex multi-stage security tasks.

\begin{table}[htbp]
\centering
\caption{The comparison of tool utilization across agent frameworks during task execution.}
\label{tab:tools_comparison}
\resizebox{0.8\linewidth}{!}{%
\begin{tabular}{l|l|c|c|c}
\hline
\textbf{Tool Type} & \textbf{Tool Capability} & \textbf{CAI} & \textbf{OpenHands} & \textbf{SWE-agent} \\
\hline
\multirow{3}{*}{\textbf{Basic Execution}} 
& execute\_bash & \checkmark & \checkmark & \checkmark \\
& str\_replace\_editor & \checkmark & \checkmark & \checkmark \\
& File Operations & \checkmark & \checkmark & \checkmark \\
\hline
\multirow{2}{*}{\textbf{Workflow Control}} 
& think (Reasoning Enhancement) & \texttimes & \checkmark & \texttimes \\
& finish (Task Summary) & \texttimes & \checkmark & \texttimes \\
\hline
\multirow{2}{*}{\textbf{Domain-Specific}} 
& Security/Vulnerability Tools & 9 & \texttimes & \texttimes \\
& Used in Tasks & 0 & N/A & N/A \\
\hline
\multirow{4}{*}{\textbf{Task Execution}} 
& Clear Task Termination & \checkmark & \checkmark & \checkmark \\
& Structured Reasoning & \texttimes & \checkmark & \texttimes \\
& Correct Artifacts Output & \texttimes & \checkmark & \texttimes \\
& Proper JSON Format & \checkmark & \checkmark & \texttimes \\
\hline
\end{tabular}%
}
\end{table}

\begin{center}
\fcolorbox{black}{gray!10}{%
\begin{minipage}{0.95\linewidth}
\textbf{Findings for C1:} Current LLM agents achieve limited web vulnerability reproduction effectiveness with end-to-end success rates below 25\%. OpenHands demonstrates superior performance through structured reasoning and systematic completion mechanisms, while tool availability alone does not ensure effectiveness. The primary bottleneck remains the exploitation gap where agents can execute PoCs but struggle to trigger actual vulnerabilities.
\end{minipage}%
}
\end{center}
\subsection{C2: Compatibility}

\paragraph{\uline{\textbf{Motivation.}}}
Real-world vulnerability reproduction environments exhibit significant diversity across vulnerability types and web technologies. Different vulnerability categories require distinct understanding patterns, exploitation techniques, and verification approaches. Similarly, vulnerabilities manifest differently across web technologies due to language-specific features, frameworks, and coding patterns. A practical vulnerability reproduction agent must demonstrate consistent performance across this diversity rather than excelling only in specific domains. Understanding compatibility helps assess whether current agents can serve as general tools or require specialized adaptations.

\paragraph{\uline{\textbf{Approach.}}}
As shown in ~\autoref{sec:dataset}, our dataset covers diverse vulnerability types and web technologies. We conduct categorical evaluation across these dimensions to assess agent compatibility. For vulnerability type compatibility, we analyze reproduction success rates across the seven vulnerability categories in our dataset. For programming language compatibility, we evaluate performance across the six primary languages. To ensure fairness and reproducibility, we use the top-performing combination from ~\autoref{sec:c1}, specifically OpenHands with Claude-Sonnet-4.

\paragraph{\uline{\textbf{Results.}}}
~\autoref{tab:vuln_type_performance} presents stage-wise success rates across different vulnerability types. The results reveal substantial performance variations, with end-to-end success rates (Success@3) ranging from 0\% to 63\%. Library-based Prototype Pollution achieves the highest reproduction success rate at 63\%. Among service-based web vulnerabilities, CSRF performs best with 60\%, followed by Path Traversal at 36\%. Other service-based vulnerabilities demonstrate considerably lower success rates: XSS at 16\%, SSRF at 15\%, RCE at 8\%, and SQL Injection with the lowest at 0\%.

The low success rates for XSS and SSRF reflect agents' difficulties with complex multi-component environments requiring database systems, frontend-backend configurations, and middleware services. Agents struggle to extract configurations from large repositories, perform data imports, and handle intricate deployment procedures. SQL Injection faces similar environment setup challenges, compounded by database-dependent architectures. Beyond environment setup, a critical bottleneck emerges in vulnerability triggering due to API authentication requirements, as agents without browser interaction capabilities cannot perform automated login procedures like human users.

In contrast, CSRF achieves perfect file localization due to clear vulnerability descriptions enabling precise interface targeting. Prototype Pollution maintains consistent performance across stages, benefiting from simpler library-based repositories requiring only dependency installation rather than complex infrastructure.

While PoC execution rates remain high across vulnerability types, the dramatic drop from execution to trigger phases reveals that agents can run exploit code but struggle with the precise environmental conditions and interaction sequences needed for actual vulnerability manifestation.
\begin{table}[htbp]
\centering
\caption{Vulnerability reproduction success rates by vulnerability types.}
\label{tab:vuln_type_performance}
\adjustbox{width=\textwidth,center}
{
\begin{tabular}{|l|c|c|c|c|c|c|c|c|}
\hline
\multirow{2}{*}{\textbf{Vulnerability Type}} & \textbf{Environment} & \multicolumn{3}{c|}{\textbf{Code Localization}} & \multicolumn{2}{c|}{\textbf{PoC Generation}} & \multicolumn{2}{c|}{\textbf{End-to-End}} \\
\cline{2-9}
 & \textbf{Setup} & \textbf{File} & \textbf{Function} & \textbf{Line} & \textbf{Execution} & \textbf{Trigger} & \textbf{Success@1} & \textbf{Success@3} \\
\hline
PROTOTYPE\_POLLUTION & 75\% & 88\% & 75\% & 63\% & 88\% & 50\% & 38\% & 63\% \\
\hline
CSRF & 80\% & 100\% & 80\% & 60\% & 80\% & 60\% & 40\% & 60\% \\
\hline
PATH\_TRAVERSAL & 45\% & 82\% & 73\% & 45\% & 82\% & 27\% & 18\% & 36\% \\
\hline
XSS & 21\% & 53\% & 37\% & 21\% & 79\% & 11\% & 5\% & 16\% \\
\hline
SSRF & 15\% & 31\% & 23\% & 8\% & 77\% & 8\% & 0\% & 15\% \\
\hline
RCE & 8\% & 17\% & 8\% & 8\% & 67\% & 8\% & 0\% & 8\% \\
\hline
SQLI & 8\% & 0\% & 0\% & 0\% & 25\% & 0\% & 0\% & 0\% \\
\hline
\textbf{TOTAL} & \textbf{29\%} & \textbf{46\%} & \textbf{36\%} & \textbf{24\%} & \textbf{70\%} & \textbf{18\%} & \textbf{10\%} & \textbf{23\%} \\
\hline
\end{tabular}
}
\end{table}

\begin{figure*}[h]
  \centering
  \includegraphics[width=0.6\linewidth]{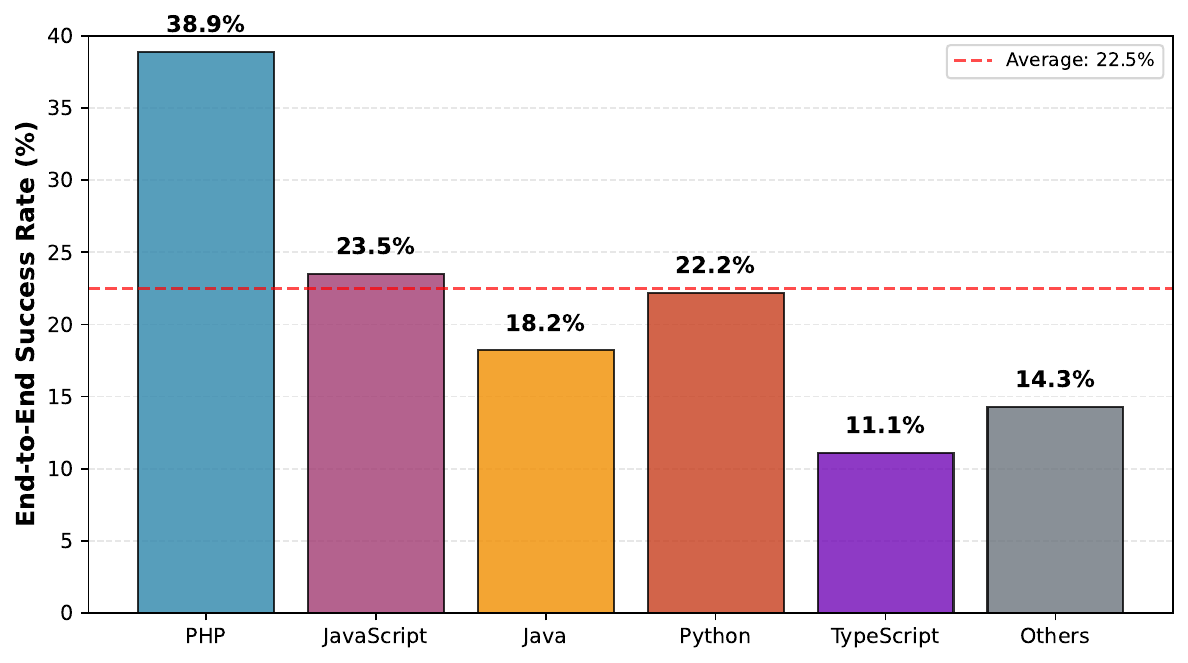}
  \caption{Comparison of vulnerability reproduction success rates between different web technologies.}
  \label{fig:language_performance}
  \Description{Overview of workflow.}
\end{figure*}

~\autoref{fig:language_performance} demonstrates end-to-end reproduction success rates across web technologies. PHP achieves the highest success rate at 38.9\%, while TypeScript shows the lowest at 11.1\%. PHP's superior performance stems from its prevalence in web security research, providing agents with abundant training data and well-documented exploit patterns, combined with its simpler syntax and direct web-oriented nature. JavaScript demonstrates strong performance through extensive client-side vulnerability examples and straightforward DOM manipulation patterns. TypeScript's poor performance shows that its additional complexity and stricter syntax requirements hinder agents despite its similarity to JavaScript. The results indicate that agent performance varies significantly across both vulnerability types and web technologies, with success rates influenced by environment complexity and the specific requirements of each vulnerability category.

\begin{center}
\fcolorbox{black}{gray!10}{%
\begin{minipage}{0.95\linewidth}
\textbf{Findings for C2:} Agent performance exhibits substantial variation across vulnerability types and web technologies. Service-based vulnerabilities face significant challenges from complex environment setup, authentication requirements, and multi-component dependencies, while library-based vulnerabilities achieve higher success through simpler deployment requirements. Language compatibility depends on security research prevalence and syntactic complexity, with web-oriented languages showing better agent adaptation.
\end{minipage}%
}
\end{center}

\subsection{C3: Efficiency}
\paragraph{\uline{\textbf{Motivation.}}}
While effectiveness measures reproduction success rates, efficiency evaluates the computational and economic costs of achieving successful reproductions. Understanding resource requirements is crucial for practical deployment, as organizations must balance reproduction capabilities with operational costs. We analyze efficiency across successful reproduction attempts to assess real-world viability of different agent-model combinations.

\paragraph{\uline{\textbf{Approach.}}}
We adopt the efficiency evaluation methodology from ~\autoref{sec:metrics}, measuring token consumption, execution time, and monetary cost across the vulnerability reproduction pipeline. To do so, we focus exclusively on the successful reproduction cases achieved by our three foundational agents as identified in ~\autoref{sec:c1}. To ensure a fair comparison, all tests are conducted using the Claude-Sonnet-4 model, and measurements are averaged across successful attempts for each agent.

\paragraph{\uline{\textbf{Results.}}}
~\autoref{tab:efficiency_results} presents efficiency metrics for all three agents calculated from successful reproduction attempts. Environment setup demonstrates the highest resource requirements across all agents, while vulnerability localization proves most economical. End-to-end reproduction costs range from \$1.68 to \$2.19 with execution times between 25.7-34.0 minutes. OpenHands consistently requires the highest resources (\$2.19 total), followed by SWE-agent (\$1.90), while CAI demonstrates the most efficient utilization (\$1.68).
\begin{table}[h]
\centering
\caption{Efficiency analysis of successful vulnerability reproduction cases.}
\label{tab:efficiency_results}
\footnotesize
\begin{tabular}{l|ccc|ccc|ccc|ccc}
\toprule
\multirow{2}{*}{\textbf{Agent}} & \multicolumn{3}{c|}{\textbf{Env}} & \multicolumn{3}{c|}{\textbf{Loc}} & \multicolumn{3}{c|}{\textbf{PoC}} & \multicolumn{3}{c}{\textbf{E2E}} \\
& \textbf{T} & \textbf{Ti} & \textbf{C} & \textbf{T} & \textbf{Ti} & \textbf{C} & \textbf{T} & \textbf{Ti} & \textbf{C} & \textbf{T} & \textbf{Ti} & \textbf{C} \\
\midrule
OpenHands & 78.3 & 18.2 & 1.17 & 15.2 & 4.3 & 0.23 & 52.7 & 11.5 & 0.79 & 146.2 & 34.0 & 2.19 \\
SWE-agent & 65.4 & 15.8 & 0.98 & 12.8 & 3.7 & 0.19 & 48.9 & 10.2 & 0.73 & 127.1 & 29.7 & 1.90 \\
CAI & 56.8 & 13.1 & 0.85 & 11.1 & 3.2 & 0.17 & 44.2 & 9.4 & 0.66 & 112.1 & 25.7 & 1.68 \\
\bottomrule
\end{tabular}
\begin{tablenotes}
\item Env: Environment Setup, Loc: Vulnerability Localization, PoC: PoC Generation, E2E: End-to-End Reproduction.
\item T: Tokens (K), Ti: Time (min), C: Cost (\$). Averages from successful cases only.
\item Costs based on Claude Sonnet pricing (\$3/1M input, \$15/1M output tokens).
\end{tablenotes}
\end{table}

The cost variations stem from architectural differences between agents. OpenHands incorporates extensive internal reasoning including explicit "think" and "finish" operations that generate substantial token overhead. SWE-agent employs numerous system prompts to enforce tool usage compliance, creating intermediate communication costs. CAI, despite having the most comprehensive tool suite, exhibits minimal tool invocation in vulnerability reproduction tasks, resulting in streamlined execution. These design choices create a trade-off between computational efficiency and reproduction capability, with OpenHands prioritizing thorough reasoning, SWE-agent emphasizing tool compliance, and CAI optimizing for direct execution.

\begin{center}
\fcolorbox{black}{gray!10}{%
\begin{minipage}{0.95\linewidth}
\textbf{Findings for C3:} Web vulnerability reproduction requires substantial computational resources, with end-to-end costs ranging from \$1.68 to \$2.19 per successful case. Environment setup consumes the most resources across all agents, while agent architectural differences create significant efficiency variations in the reproduction pipeline.\end{minipage}%
}
\end{center}

\subsection{C4: Robustness Analysis}

\paragraph{\uline{\textbf{Motivation.}}}
Robustness measures how well agents perform under challenging conditions. In vulnerability reproduction, agents face obstacles like limited processing capabilities, browser security restrictions, or incomplete information. These challenges are inherent to real-world security research environments where perfect conditions rarely exist. A robust agent should maintain reasonable performance across different scenarios, adapting to unexpected failures and environmental constraints. Understanding robustness helps us assess agent reliability for practical deployment and identify which architectures handle reproduction challenges most effectively.

\paragraph{\uline{\textbf{Approach.}}}
We design two experiments to test the robustness of our three selected agents across key challenge areas: First, we compare agent performance using Claude-Sonnet-4 with and without enhanced reasoning across different vulnerability reproduction stages in 10 vulnerabilities. Second, we test agent performance under different authentication scenarios using 12 SQL Injection vulnerabilities, comparing three cases: manually providing login tokens, prompting agents to login themselves, and providing no authentication help.

\begin{figure}[htbp]
  \centering
  \includegraphics[width=0.8\linewidth]{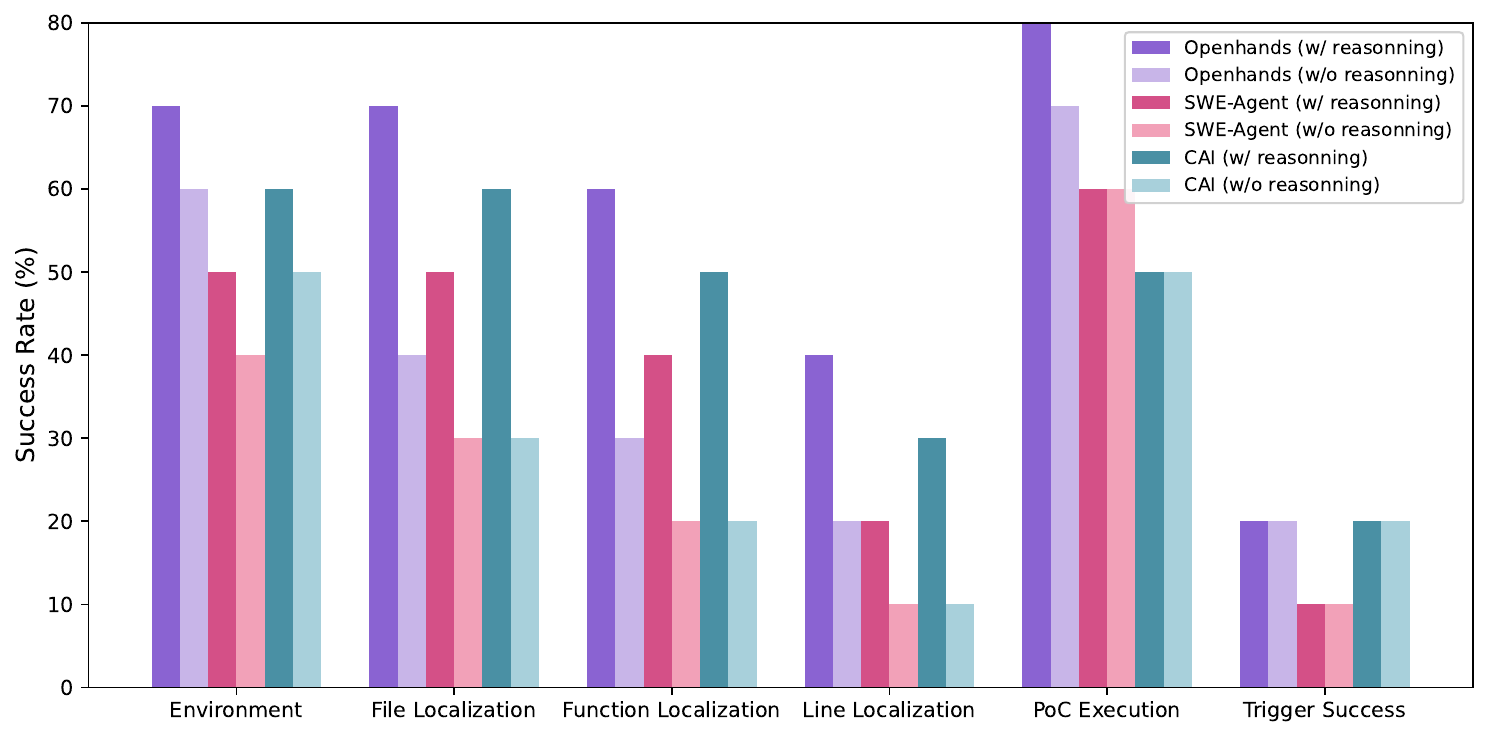}
  \caption{Performance comparison between agents using Claude-Sonnet-4 with and without reasoning across different vulnerability reproduction stages.}
  \label{fig:reasoning}
\end{figure}

\paragraph{\uline{\textbf{Results.}}}
~\autoref{fig:reasoning} shows mixed results when comparing models with and without enhanced reasoning. Enhanced reasoning helps in early stages, with OpenHands improving from 60\% to 70\% in Environment Setup and from 40\% to 70\% in File Localization, while improvements decrease in later stages with PoC Execution showing smaller gaps and Trigger Success showing no improvement across all agents. SWE-Agent shows the most consistent benefit from enhanced reasoning across localization stages, while OpenHands demonstrates strong baseline performance even without reasoning due to its built-in "think" operations, and CAI shows more variable results with minimal reasoning benefits. The limited reasoning impact reflects both agent design differences and task complexity variations, as reasoning enhancements prove most effective for analytical tasks like code localization but provide little benefit for execution-oriented activities like vulnerability triggering. Within our 10-vulnerability sample, agents demonstrate robustness across both reasoning configurations, suggesting that architectural design choices may be more influential than underlying model reasoning capabilities for practical vulnerability reproduction tasks.

\autoref{tab:auth-robustness} shows authentication context impacts vulnerability reproduction success. Manual token provision works best, with Openhands + Claude-Sonnet-4 achieving 67\% success. Requiring agents to handle login autonomously results in an average performance degradation of 33.3\% compared to manual token provision, with success rates dropping to 25-50\% across agent-model combinations. Removing authentication guidance causes near-complete failure, with only three configurations maintaining 8\% success rates. This reveals that current agents struggle with autonomous authentication discovery and rely heavily on explicit guidance. The consistent degradation across architectures and models indicates that authentication handling is a widespread limitation.
\begin{table}[htbp]
\centering
\caption{Comparison of trigger rate in authentication robustness analysis.}
\label{tab:auth-robustness}
\resizebox{0.7\linewidth}{!}{%
\begin{tabular}{llccc}
\toprule
\textbf{Agent} & \textbf{Model} & \textbf{Manual Token} & \textbf{Agent Login} & \textbf{No Auth Info} \\
\midrule
\multirow{3}{*}{Openhands} & Claude-Sonnet-4 & 67\% & 50\% & 8\% \\
 & Gemini-2.5-Pro & 58\% & 42\% & 8\% \\
 & GPT-4.1 & 50\% & 33\% & 0\% \\
\midrule
\multirow{3}{*}{SWE-Agent} & Claude-Sonnet-4 & 58\% & 33\% & 8\% \\
 & Gemini-2.5-Pro & 50\% & 25\% & 0\% \\
 & GPT-4.1 & 42\% & 17\% & 0\% \\
\midrule
\multirow{3}{*}{CAI} & Claude-Sonnet-4 & 50\% & 25\% & 0\% \\
 & Gemini-2.5-Pro & 42\% & 17\% & 0\% \\
 & GPT-4.1 & 33\% & 8\% & 0\% \\
\bottomrule
\end{tabular}%
}
\end{table}


\begin{center}
\fcolorbox{black}{gray!10}{%
\begin{minipage}{0.95\linewidth}
\textbf{Findings for C4:} Current agents show robustness to reasoning changes but high sensitivity to prompt variations. While reasoning versus non-reasoning modes produce similar performance, incomplete authentication results in severe performance drops, highlighting agents' limited capacity for problem-solving without explicit guidance.
\end{minipage}%
}
\end{center}

\section{Discussion}
\label{sec:discussion}

Our empirical evaluation reveals that current open-source agents face substantial limitations in vulnerability reproduction beyond specific ecosystems. While agents demonstrate reasonable success in library-based JavaScript environments and relatively straightforward CSRF scenarios, they struggle significantly with complex web vulnerabilities that require intricate environment setup and multi-step exploitation chains. The stark performance gaps—ranging from 0\% for SQL injection to 63\% for prototype pollution—indicate that pure code-based approaches are insufficient for comprehensive vulnerability reproduction. Traditional injection vulnerabilities, despite their well-documented nature, consistently fail due to environment configuration challenges and the complex interplay between application deployment, database setup, and exploitation context. This suggests that current agent architectures fundamentally lack the systems-level understanding necessary for realistic security testing scenarios.

The integration of Model Context Protocol (MCP)\cite{mcp_specification_2025} with browser automation tools like Playwright or Chrome presents a breakthrough for vulnerability reproduction. By enabling agents to control browser instances, capture network traffic, and monitor application state changes in MCP-based approaches can bridge the gap between static code analysis and dynamic exploitation. This browser-centric methodology allows agents to observe runtime behavior of web applications, intercept HTTP requests and responses, and validate success through DOM manipulation and state inspection. The ability to interact with web interfaces while monitoring network activity represents a shift from traditional code-only approaches, unlocking automated reproduction capabilities for complex web vulnerabilities that have resisted agent-based solutions.
\section{Threats to Validity}
\label{sec:threats}

\textbf{Internal Validity.} The primary internal threat stems from potential data leakage in LLM training datasets. Many CVEs in our benchmark were disclosed before the training cutoff dates of evaluated models, creating a risk that agents may leverage memorized exploit patterns rather than demonstrating genuine reasoning capabilities. Although we deliberately avoided referencing specific PoCs in our prompts, models may still access historical vulnerability data through their pre-trained knowledge, potentially inflating performance metrics. Additionally, our evaluation employs binary success metrics that only capture whether vulnerability reproduction succeeds or fails, providing a single-dimensional assessment that may miss important aspects of the reproduction process.

\noindent\textbf{External Validity.} Our findings may have limited generalizability due to several factors. First, our CVE selection, while diverse across vulnerability types and web technologies, represents a curated subset that may not fully reflect the complexity distribution of real-world security scenarios. Second, the containerized evaluation environment, despite efforts toward realism, may still differ from production deployments in ways that affect agent performance. Finally, the rapid evolution of LLM capabilities means our findings represent a temporal snapshot that may not predict future agent performance across different models or architectures.
\section{Related Work}
\label{sec:related_work}
\subsection{Software Engineering Benchmarks for LLM Agents}
Software engineering represents a critical application domain for evaluating LLM agents, with benchmarks spanning different complexity levels and task types. SWE-bench~\cite{jimenez2024swe} evaluates agents' abilities to solve real-world GitHub issues by generating patches for bug fixes, representing one of the most challenging repository-level tasks that require understanding complex codebases and coordinating changes across multiple files. Traditional coding benchmarks like HumanEval focus on isolated function generation from natural language descriptions, while MBPP evaluates basic Python programming tasks. Additional benchmarks including BigCodeBench~\cite{zhuo2024bigcodebench}, LiveCodeBench~\cite{jain2024livecodebench}, and EvalPlus~\cite{liu2023evalplus} extend evaluation across multiple languages and enhanced test coverage. However, these SE benchmarks exhibit significant limitations: many focus on narrow task scopes rather than diverse professional tasks, traditional benchmarks often involve only local edits to individual functions rather than repository-wide reasoning, and they frequently ignore autonomous environment setup and end-to-end task solving capabilities essential for practical software engineering. Recent advances like AgentBench~\cite{liu2023agentbench} provide more comprehensive evaluation across diverse environments, highlighting the evolution toward more realistic and challenging evaluation frameworks for software engineering agents.

\subsection{Cybersecurity Benchmarks for LLM Agents}
Various specialized benchmarks have been developed to evaluate LLM agents in cybersecurity. InterCode-CTF~\cite{yang2023language} manually evaluated LLMs on Capture-the-Flag challenges requiring reverse engineering and exploitation skills, while NYU CTF Bench~\cite{shao2024nyuctfbench} expanded this with an automated framework for diverse offensive security tasks across cryptography, forensics and binary exploitation. The CyberSecEval series~\cite{bhatt2024cyberseceval2,wan2024cyberseceval3} focuses on evaluating cybersecurity risks including prompt injection, code interpreter abuse and automated social engineering. ARVO~\cite{mei2024arvo} provides reproducible memory vulnerabilities with triggering inputs and developer patches. SEC-bench~\cite{lee2025secbench} evaluates PoC generation and vulnerability patching tasks using data from OSS-FUZZ project, SecVulEval~\cite{ahmed2025secvuleval} benchmarks statement-level vulnerability detection in C/C++ code. CyberGym~\cite{wang2025cybergym} assesses vulnerability reproduction across large software projects. Most existing benchmarks exhibit limitations in capturing real-world cybersecurity complexity, often focusing on isolated tasks or synthetic environments. The most closely related work is CVE-bench~\cite{wang2025cvebench}, which remains limited in both scope and scale, capturing only a narrow slice of agent performance in web vulnerability reproduction scenarios.

\section{Conclusion}
\label{sec:conclusion}

Our evaluation of 20 LLM agents across 16 core dimensions on 3 representative CVEs identified three top performers (OpenHands, SWE-agent, and CAI) for testing on 80 real-world CVEs spanning 6 web technologies and 7 vulnerability types. Results show agents handle simple library-based vulnerabilities well but consistently fail on complex service-based scenarios requiring multi-component environments and authentication. The gap between executing exploit code and triggering actual vulnerabilities reveals fundamental limitations in environmental adaptation.
These findings highlight the need for significant advances before automated vulnerability reproduction becomes practical. Current agents lack the robustness required for real-world deployment, showing high sensitivity to input guidance and poor performance under incomplete information.

\section*{Data Availability}

The artifact of this paper can be accessed via \url{https://figshare.com/s/7e55eaaaca0b0146ee62}.

\bibliographystyle{ACM-Reference-Format}
\bibliography{main}

\end{document}